\documentclass[10pt,aps,prd,reprint,superscriptaddress,nofootinbib,preprintnumbers]{revtex4-1}

\usepackage{amsmath}
\usepackage{graphicx}
\usepackage[colorlinks,pdfusetitle]{hyperref}
\hypersetup{colorlinks=true,allcolors=[rgb]{1,0.56,0}}

\newcommand{\orcid}[1]{\href{https://orcid.org/#1}{#1}}
\newcommand{\wh}[1]{\widehat{#1}}
\newcommand{\Dmsqee}{\Delta m^2_{ee}}
\newcommand{\blue}[1]{\color{blue} #1 \color{black}}

\begin{document}

\preprint{FERMILAB-PUB-24-0013-T}

\title{The Smallness of Matter Effects in Long-Baseline Muon Neutrino Disappearance}

\author{Peter B.~Denton}
\email{pdenton@bnl.gov}
\thanks{\orcid{0000-0002-5209-872X}}
\affiliation{High Energy Theory Group, Physics Department, Brookhaven National Laboratory, Upton, NY 11973, USA}

\author{Stephen J.~Parke}
\email{parke@fnal.gov}
\thanks{\orcid{0000-0003-2028-6782}}
\affiliation{Theoretical Physics Department, Fermi National Accelerator Laboratory, P.~O.~Box 500, Batavia, IL 60510, USA}

\date{March 7, 2024}

\begin{abstract}
Current long-baseline accelerator experiments, NOvA and T2K, are making excellent measurements of neutrino oscillations and the next generation of experiments, DUNE and HK, will make measurements at the $\mathcal O(1\%)$ level of precision.
These measurements are a combination of the appearance channel which is more challenging experimentally but depends on many oscillation parameters, and the disappearance channel which is somewhat easier and allows for precision measurements of the atmospheric mass splitting and the atmospheric mixing angle.
It is widely recognized that the matter effect plays a key role in the appearance probability, yet the effect on the disappearance probability is surprisingly small for these experiments.
Here we investigate both exactly how small the effect is and show that it just begins to become relevant in the high statistics regime of DUNE.
 
\vspace*{0.1in}
 
\noindent
\blue{
\emph{We dedicate this paper to the memory of the outstanding neutrino theorist, inspiring mentor, and dear friend, \href{https://inspirehep.net/authors/983874}{Tom Weiler}.
His insights on neutrino physics and the neutrinos he emitted will both live on until the end of the Universe.}
}
\end{abstract}

\maketitle

\section{Introduction}
Considerable progress has been made in determining the six oscillation parameters.
Addressing the three remaining unknowns, the sign of the atmospheric mass splitting, the octant of $\theta_{23}$, and measuring the amount of CP violation, requires a high degree of precision of all the other oscillation parameters \cite{Denton:2022een,Denton:2023zwa}.
For the absolute value of the atmospheric mass splitting and the closeness of $\theta_{23}$ to maximal, the best approach is via $\nu_\mu$ disappearance in long-baseline experiments with additional information in the same oscillation channel with atmospheric neutrinos.

The matter effect \cite{Wolfenstein:1977ue} modifies the oscillation probability in the presence of matter and is relevant for long-baseline experiments through the Earth's crust.
In particular, it modifies the propagation of $\nu_e$ and $\bar\nu_e$'s.
For the appearance channels, $\nu_\mu\to\nu_e$ and $\bar\nu_\mu\to\bar\nu_e$, the impact is sizable, see e.g.~\cite{Barger:1980tf,Zaglauer:1988gz,Arafune:1997hd,Cervera:2000kp,Freund:2001pn,Akhmedov:2004ny,Friedland:2006pi,Asano:2011nj,Agarwalla:2013tza,Minakata:2015gra,Denton:2016wmg,Kelly:2018kmb,Barenboim:2019pfp} and plays a major role in DUNE's sizable sensitivity to the atmospheric mass ordering \cite{DUNE:2020ypp}.

The matter effect also plays a role in $\nu_\mu$ disappearance, but the effect is typically smaller; understanding this is the focus of this article.
While it is true that for $\nu_\mu$ disappearance experiments the neutrinos spend the majority of their time in a state that would project on to either $\nu_\mu$ or $\nu_\tau$, they still spend $\mathcal O($few$\%)$ of the time on average in a $\nu_e$ state.
This alone does not explain the smallness of the impact of the matter effect which is at the $\mathcal O(0.1\%)$ level or less as shown in this paper.
This level is below the expected experimental determination of $\Delta m^2_{31}$ which will be at the $0.5-1\%$ level in the upcoming long-baseline accelerator neutrino experiments.

In this paper we will work in a framework where we take the vacuum oscillation probability (either two flavor or three flavor) and modify the six oscillation parameters as they evolve in the presence of matter.
Specifically, we will write probabilities as a function of the eigenvalues and eigenvectors (mapped onto mixing angles using the standard parameterization \cite{ParticleDataGroup:2022pth,Denton:2020igp} of the PMNS matrix \cite{Pontecorvo:1957cp,Maki:1962mu}).
This makes a determination of the evolution of the mixing parameters directly related to the impact of the matter effect.
Then we will investigate the role of the matter effect on both the $\Delta m^2$ and the mixing angles.

The article is organized as follows.
We will overview the role of the matter effect in neutrino oscillations in general and a basic understanding of $\nu_\mu$ disappearance in vacuum in section \ref{sec:overview}.
We then present numerical results to quantify the size of the effects discussed in section \ref{sec:numerical}.
In section \ref{sec:evolution delta msq} we will present our main results of the evolution of the relevant $\Delta m^2$ in matter and the mixing angles (e.g.~amplitude of the oscillation) in matter.
Finally we will discuss our results in a broader context including atmospheric neutrinos in section \ref{sec:discussion} and conclude in section \ref{sec:conclusions}.

\section{\texorpdfstring{$\nu_\mu$}{nu mu} Disappearance and the Matter Effect}
\label{sec:overview}
Neutrino oscillation probabilities are expressed as a function of the eigenvalues and eigenvectors of the Hamiltonian.
In vacuum, the probabilities are thus directly a function of the fundamental parameters: the neutrino masses (expressed as $\Delta m^2_{ij}\equiv m_i^2-m_j^2$) and parameters of the mixing matrix between the mass basis and the flavor or interaction basis.
The Hamiltonian in vacuum is
\begin{equation}
H=\frac1{2E}\left[U
\begin{pmatrix}
0\\&\Delta m^2_{21}\\&&\Delta m^2_{31}
\end{pmatrix}U^\dagger\right]\,,
\end{equation}
where $E$ is the neutrino energy and the mixing matrix is a function of four degrees of freedom for oscillations and can be parameterized in a variety of ways \cite{Denton:2020igp}.
It is usually parameterized as
\begin{equation}
U\hspace{-1mm}=\hspace{-1mm}
\begin{pmatrix}
1\\&c_{23}&s_{23}\\&-s_{23}&c_{23}
\end{pmatrix}
\hspace{-2mm}
\begin{pmatrix}
c_{13}&&s_{13}e^{-i\delta}\\&1\\-s_{13}e^{i\delta}&&c_{13}
\end{pmatrix}
\hspace{-2mm}
\begin{pmatrix}
c_{12}&s_{12}\\-s_{12}&c_{12}\\&&1
\end{pmatrix}
\hspace{-1mm},
\end{equation}
where $s_{ij}=\sin\theta_{ij}$ and $c_{ij}=\cos\theta_{ij}$.
For antineutrinos shift $\delta\to-\delta$.

The muon neutrino disappearance probability in vacuum is given by
\begin{align}
P_{\mu\mu}^{\rm vac} &=1- \sum_{i>j}4 |U_{\mu i} |^2 |U_{\mu j} |^2  \sin^2\frac{\Delta m^2_{ij}L}{4E} \,.
\end{align}
It has been known that, given the oscillation parameters, $\nu_\mu$ disappearance in vacuum is well described in an effective two-flavor picture near the first oscillation minimum.
This probability can be written as
\begin{equation}
P_{\mu\mu}^{\rm vac}\simeq1-\sin^22\theta_{23}\sin^2\frac{\Delta m^2_{\mu\mu,{\rm vac}}L}{4E}\,,
\end{equation}
where the best approximation for the frequency \cite{Nunokawa:2005nx} is
\begin{align}
\Delta m^2_{\mu\mu,{\rm vac}}\equiv{}&m_3^2-\frac{|U_{\mu1}|^2m_1^2+|U_{\mu2}|^2m_2^2}{|U_{\mu1}|^2+|U_{\mu2}|^2}\,,\label{eq:Dmsqmm effective}\\
={}& \frac{m_3^2 - m_\mu^2}{1-|U_{\mu3}|^2}  \quad  \text{with} ~m_\mu^2 \equiv  \sum m_i^2 |U_{\mu i}|^2 \notag \,, \\[2mm]
\approx{}&s_{12}^2\Delta m^2_{31}+c_{12}^2\Delta m^2_{32}\nonumber\\
&+s_{13}\sin2\theta_{12}\tan\theta_{23}\cos\delta\Delta m^2_{21}\,,\label{eq:Dmsqmm effective approx}
\end{align}
where the approximation in the third equation drops terms proportional to $s_{13}^2$.
The term $m_\mu^2$ is defined to be equal to $2EH_{\mu\mu}$ where $H_{\mu\mu}$ is the $\mu,\mu$ term of the Hamiltonian, see appendix \ref{sec:hamiltonian}.
The final expression above is useful in many situations, but the middle expression, which is slightly more accurate, will prove most useful in the discussion of the matter effect.

The matter effect induces an additional potential that modifies the $\nu_e$ energy.
Specifically, $H_{ee}$ (the top left element of the Hamiltonian) gains an additional term that is $a/2E$ where $a=2\sqrt2EG_FN_e$ quantifies the size of the matter potential, $G_F$ is Fermi's constant, and $N_e$ is the average electron density along the propagation.
The focus of this article is to understand when a two-flavor approximation is valid even in the presence of the matter effect.

\section{Numerical Study}
\label{sec:numerical}
In this section we perform a series of numerical studies examining the role of the matter effect on $\nu_\mu$ disappearance, focused on the first oscillation minimum, to further motivate the importance of our analytic studies below.

\begin{figure}
\centering
\includegraphics[width=\columnwidth]{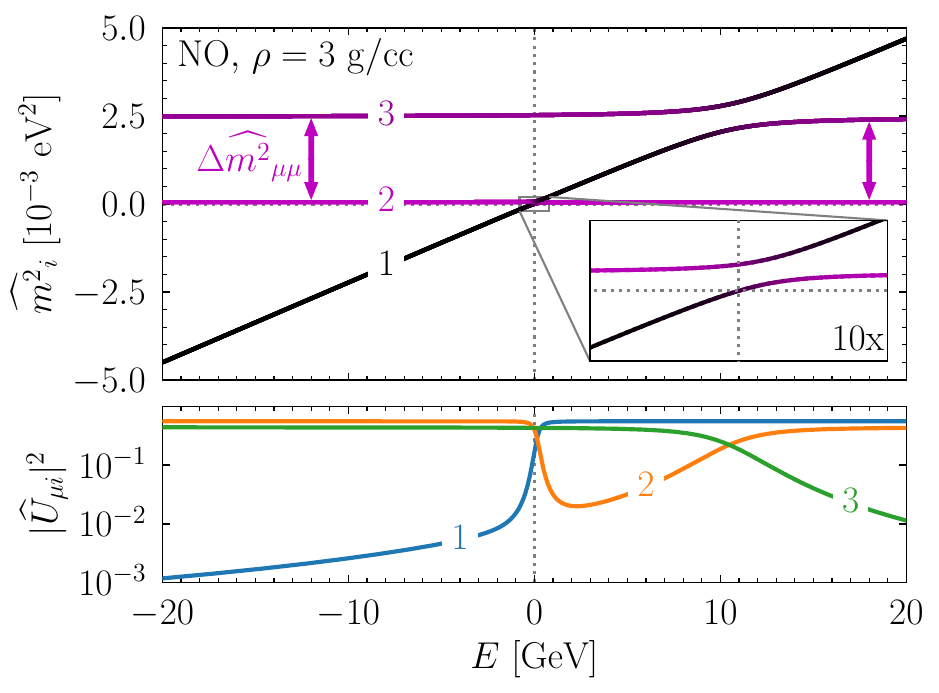}
\caption{\textbf{Top}: The three eigenvalues in matter, $\widehat{m^2_i}$,  colored by their $\nu_\mu$ weighted fraction; the purple sections are more $\nu_\mu$ and black sections have little $\nu_\mu$ in them.
The vertical arrows designate the effective $\Delta \widehat{m^2}$ at that energy.
We have taken the density to be 3 g/cc and the mass ordering normal.
The inset shows the color change around the solar resonance.
\textbf{Bottom}: The muon content of each of the three eigenvectors in matter: $|\widehat{U}_{\mu i}|^2$ which provide the coloring for the top panel.
Antineutrinos are denoted by negative energies.}
\label{fig:eigenvalues}
\end{figure}

First, we calculate the exact eigenvalues and eigenvectors in matter for the normal ordering (NO) at the Earth's crust density of $\rho=3$ g/cc in fig.~\ref{fig:eigenvalues}.
We highlight in purple the relative fraction of each eigenvalue that is muon neutrino, see also the lower panel for the norm square of each component of the muon neutrino eigenvector.
We see why below the atmospheric resonance at about 11 GeV for neutrinos (in the inverted ordering (IO) the resonance is for antineutrinos) the separation between the two dominantly $\nu_\mu$ eigenvalues is relatively constant, suggesting that the actual effect $\Delta m^2$ may not depend much on neutrino energy below the atmospheric resonance.

\begin{figure*}
\centering
\includegraphics[width=\textwidth]{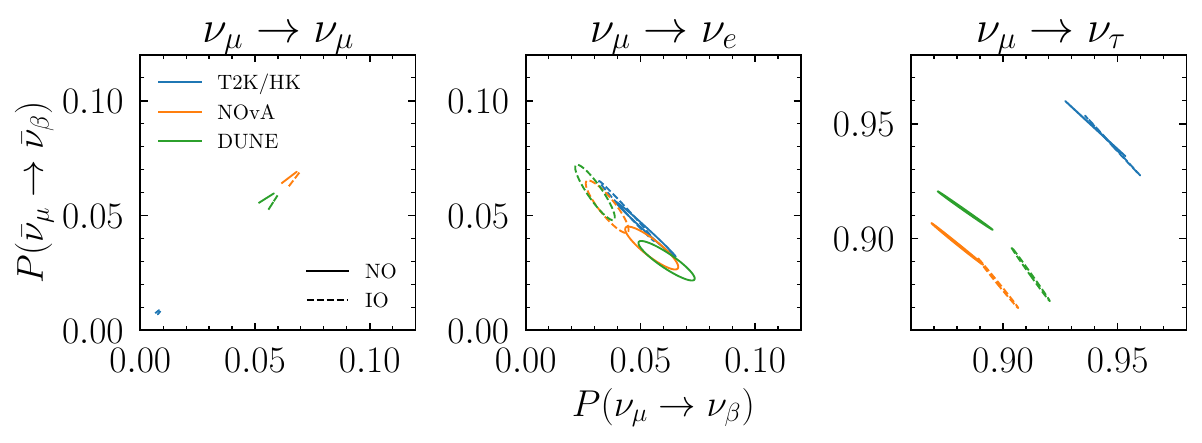}
\caption{The probability in neutrino mode (horizontal axis) and antineutrino mode (vertical axis) for the three oscillation channels from muon neutrinos.
The disappearance probabilities, $\nu_\mu\to\nu_\mu$ are shown on the \textbf{left}, while the appearance probabilities are shown in the \textbf{center} ($\nu_\mu\to\nu_e$) and the \textbf{right} ($\nu_\mu\to\nu_\tau$) right panels.
The two mass orderings are denoted by solid curves for the NO and dashed curves for the IO.
The curves themselves denote changing $\delta$ around $[0,2\pi)$.
Three different experiments are shown, T2K/HK (blue), NOvA (orange), and DUNE (green) at energies of 0.6, 1.9 and 3.0 GeV, respectively, see table \ref{tab:exps} for further details. Here, when switching between mass orderings we hold $|\Delta m^2_{32}+s^2_{12}\Delta m^2_{21}|$ fixed. In appendix \ref{biprobabilities} we discuss more details on the features of these figures.}
\label{fig:biprobability}
\end{figure*}

Second, in fig.~\ref{fig:biprobability}, we plot the probability in both the neutrino and antineutrino modes for the three oscillation channels for accelerator neutrinos: first the disappearance case $\nu_\mu\to\nu_\mu$, then the two appearance cases: $\nu_\mu\to\nu_e$ and $\nu_\mu\to\nu_\tau$.
We show the standard biprobability ellipses by varying delta for not only the $\nu_\mu\to\nu_e$ channel, but also the other two, as well as both mass orderings.
We consider three different experimental setups as described in table \ref{tab:exps}.
The energies we have chosen represent the peak event rate; while small changes in the energy can lead to modest changes in the shapes of the ellipses, the overall narrative does not change.
We see that in the disappearance channel there is a small, but non-zero difference between the two mass orderings indicating a small but nonzero dependence on the matter effect.
The effect is largest for DUNE and smallest for T2K/HK.
This is because in vacuum it is impossible to differentiate between the mass orderings.
We also see that since the role of the matter effect is small for $\nu_\mu\to\nu_\mu$ disappearance, $\nu_\mu\to\nu_\tau$ disappearance must have a fairly large dependence on the matter effect since it is well known that $\nu_\mu\to\nu_e$ depends on the matter effect quite a bit, combined with the unitarity statement: $\sum_\alpha P_{\mu\alpha}=1$.

\begin{table}
\centering
\caption{The standard values used throughout the text for the long-baseline accelerator experiment configurations.}
\label{tab:exps}
\vspace*{2mm}
\begin{tabular}{c|c|c|c}
Exp&$L$ [km]&$\rho$ [g/cc]&$E$ [GeV]\\\hline
T2K/HK&295&2.6&0.60\\
NOvA&810&2.84&1.9\\
DUNE&1300&2.84&3.0
\end{tabular}
\end{table}

\begin{figure}
\centering
\includegraphics[width=\columnwidth]{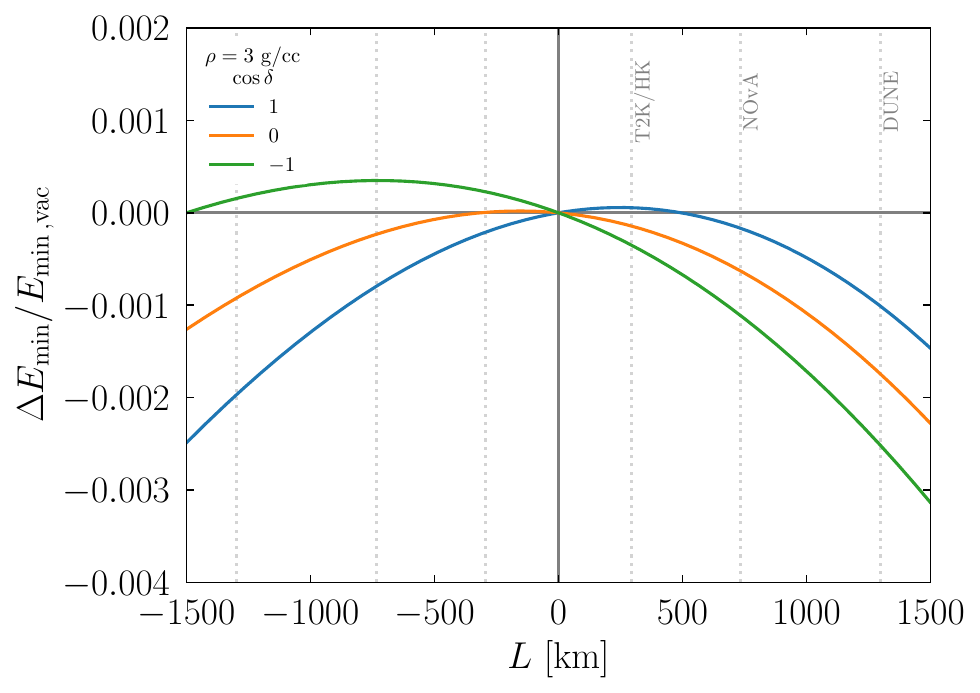}
\caption{The fractional change in the energy of the first oscillation minimum compared to the value in vacuum for experiments at different baselines in the NO for different values of $\cos\delta$; the IO is similar with neutrino and antineutrino interchanged.
Baselines for existing and upcoming accelerator experiments are denoted.
Antineutrinos are shown on the left side of the plot with $L<0$.}
\label{fig:Emin vs L}
\end{figure}

Third, we numerically evaluate the oscillation minimum in matter as shown in fig.~\ref{fig:Emin vs L} for different values of $\cos\delta$.
That is, given a long-baseline accelerator disappearance experiment at some baseline $L$, we numerically evaluate the energy at the first oscillation minimum $E_{\min}$ and compare to the energy at the minimum in vacuum.
The fractional deviation in $E_{\min}$ relative to the vacuum value is shown to vary by $<0.3\%$ across all long-baseline accelerator experiments.
We also show the approximate expression for the vacuum from eq.~\ref{eq:Dmsqmm effective}.

While fig.~\ref{fig:Emin vs L} shows the small variation in the energy at the oscillation minimum, and thus the effective value of $\Delta m^2_{\mu\mu}$, in a real experiment the statistics are very low there, and it could be the case that the matter effect plays a larger role away from the minimum.
Thus we also investigate if there is any impact in the shape of the oscillation probability away from the minimum due to the presence of matter.

\begin{figure}
\centering
\includegraphics[width=\columnwidth]{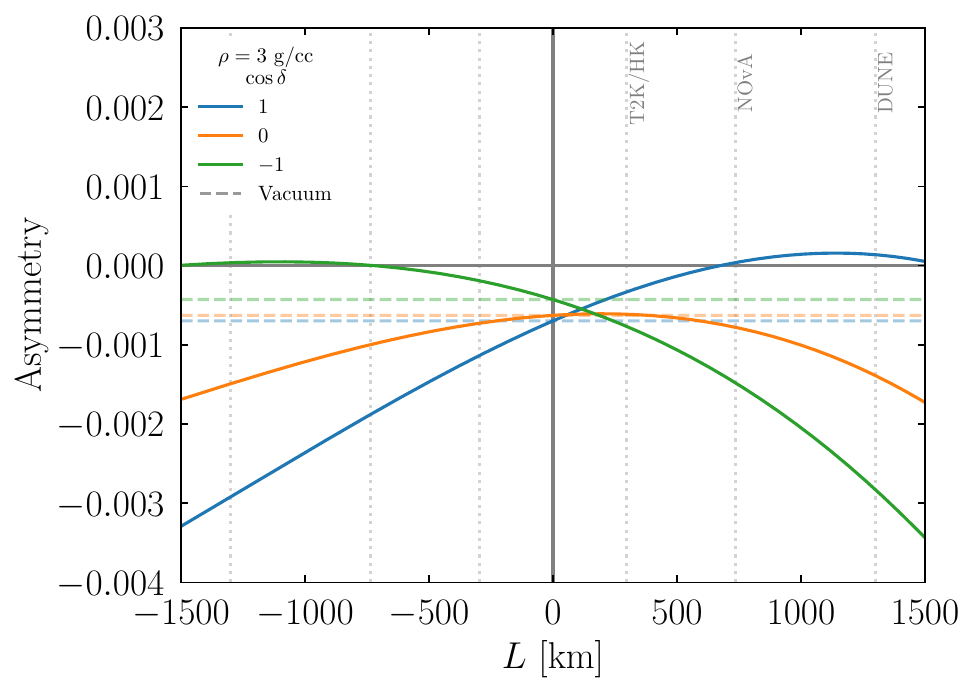}
\caption{The asymmetry (see eq.~\ref{eq:asymmetry} in the text) around the oscillation minimum as a function of the baseline and $\cos\delta$.
The asymmetries for vacuum oscillations are highlighted with the dashed curve.
Antineutrinos are shown on the left side of the plot with $L<0$.}
\label{fig:asymmetry}
\end{figure}

The probability will be symmetric in inverse energy around the minimum for two-flavor oscillations in vacuum, and we use this starting point to quantify any shape effects.
We note that the shape will not be exactly as simple as the vacuum two-flavor probability because 1) the solar $\Delta m^2_{21}$ contribution to the probability will cause a slight asymmetry and 2) both the $\Delta m^2$'s and the mixing angles depend on the neutrino energy via the matter effect.
To quantify this shift we introduce several new variables since it is convenient to work in inverse energy space.
We define $x=E^{-1}$ and then $x_{\min}$ at the first oscillation minimum.
We then $x_U$ and $x_L$ such that $x_U>x_{\min}$ and $x_L<x_{\min}$ and $P(x_L)=P(x_U)=(1+P_{\min})/2$ (note that defining $x_{L,U}$ via something like $P=1/2$ here leads to no significant change).
That is, the $x_{L,U}$ are defined as the inverse energies where the probabilities are halfway between the highest probability, 1, and the probability at the oscillation minimum, $P_{\min}$, which is close to zero.
Then we define the asymmetry as
\begin{align}
{\cal A}&\equiv\frac{(x_U-x_{\min})-(x_{\min}-x_L)}{(x_U-x_{\min})+(x_{\min}-x_L)}\,,\label{eq:asymmetry}\\
&=\frac{(x_U-x_{\min})-(x_{\min}-x_L)}{x_U-x_L}\,.
\end{align}
Thus ${\cal A}=0$ for a two-flavor vacuum picture.
We show the asymmetry around the first oscillation minimum of $\nu_\mu$ disappearance as a function of baseline in fig.~\ref{fig:asymmetry} and see that while the effect generally increases with baseline and varies somewhat with $\cos\delta$, it is also $<0.3\%$ for DUNE and smaller for other long-baseline experiments.
We also see that both the matter effect does contribute to the effect for DUNE and the deviations from the two-flavor play an important role for DUNE and, to a lesser extent, NOvA.
The asymmetry for T2K/HK is dominated by the three-flavor effect rather than the matter effect.
Thus we conclude that while shape effects could be used to probe the three-flavor picture or the matter effect in muon neutrino disappearance, in practice the effect is very small.

\section{Constructing an Accurate Two-Flavor Approximation}
\label{sec:evolution delta msq}
We now explain the smallness of the role of the matter affect by analytically studying the impact of the matter effect on the oscillation parameters, both the $\Delta m^2$ and the mixing angles by continuing to approximate the $\nu_\mu$ disappearance channel in matter as a two-flavor picture.
We will show both that this is a good approximation and that the impact of the matter effect is small, consistent with the numerical results above.
We begin by focusing on the role of the $\Delta m^2$.

We begin by taking the vacuum expression in eq.~\ref{eq:Dmsqmm effective} and elevating the parameters to matter variables,
\begin{equation}
\wh{\Delta m^2}_{\mu\mu}\approx\frac{\widehat{m^2_3}-m^2_\mu}{1-|\wh U_{\mu3}|^2}\,,\label{eq:Dmsqmm eigenvalues}
\end{equation}
where $\wh{m^2_3}$ is the third eigenvalue in matter, $\wh U_{\alpha i}(a)$ is the matrix that diagonalizes the Hamiltonian in matter, and
\begin{equation}
m^2_\mu=\sum_i|\wh U_{\mu i}|^2\wh{m^2_i}=\sum_i|  U_{\mu i}|^2 m^2_i=2EH_{\mu\mu}\,,
\end{equation}
and is the same as in vacuum; see appendix \ref{sec:hamiltonian} for the $|U_{\alpha i}|^2$ elements and the elements of the Hamiltonian in matter written out explicitly in the flavor basis.

Note that taking the approximations of eqs.~\ref{eq:Dmsqmm effective} and \ref{eq:Dmsqmm effective approx} and elevating those parameters to their matter equivalents gives an approximation that is sufficiently different that it is important to use the full expression.
This is because, as we have seen, we are investigating effects $\mathcal O(0.3\%)$ or less and the terms dropped for that approximation are $\mathcal O(s_{13}^2\simeq2\%)$.

Then, for an experiment at a given $L$, we can approximate the energy at the oscillation minimum by solving
\begin{equation}
\frac{\wh{\Delta m^2_{\mu\mu}}L}{4E}=\frac\pi2\,,
\label{eq:Dmsq pi2}
\end{equation}
for $E=E_{\min}$.

This approximation is somewhat accurate and explains some of the impact of the matter effect, but it is clear that more information is needed as the fractional error is comparable to the size of the effect.

\vspace*{0.1in}

Next, we include the fact that the amplitude of the oscillations also depends on the matter effect and thus the energy and must also be included.
We write our full two-flavor approximation as
\begin{equation}
P_{\mu\mu}\approx1-4|\wh U_{\mu3}|^2(1-|\wh U_{\mu3}|^2)\sin^2\frac{\wh{\Delta m^2_{\mu\mu}}L}{4E}\,.
\label{eq:NPZ probability}
\end{equation}
To find the energy at the minimum we now must minimize this probability numerically.

The components of the eigenvectors in matter can be easily expressed as a function of the eigenvalues and elements of the Hamiltonian via the ``Rosetta'' expression \cite{Denton:2019ovn}.
After some rearranging we find
\begin{equation}
|\wh U_{\mu3}|^2=\frac{(\wh{m^2_3})^2-\mathcal S\wh{m^2_3}+\mathcal P}
{3(\wh{m^2_3})^2-2A\wh{m^2_3}+B}\,,
\label{eq:Um3msq}
\end{equation}
where the sum ($\mathcal S$) and product ($\mathcal P$) terms are simple functions of the Hamiltonian in the flavor basis.
Using the convention that $(m^2_1,m^2_2,m^2_3)=(0, \Delta m^2_{21}, \Delta m^2_{31})$ then
\begin{align}
\mathcal S={}&2E\,(H_{ee}+H_{\tau\tau} )  \\
={}& \Delta m^2_{21}(1-|U_{\mu 2}|^2)+ \Delta m^2_{31}(1-|U_{\mu3}|^2)+a \,, \notag \\
\mathcal P={}&4E^2\,(H_{ee}H_{\tau\tau}-|H_{e\tau}|^2)\,,  \\
={}&  \Delta m^2_{21} \Delta m^2_{31}|U_{\mu1}|^2  
+a( \Delta m^2_{21}|U_{\tau 2}|^2+ \Delta m^2_{31}|U_{\tau3}|^2) \,.  \notag
\end{align}
where the dependence on the matter potential only comes from $H_{ee}$ and is linear.
The other variables in eq.~\ref{eq:Um3msq}, $A$ and $B$, come from the characteristic equation
\begin{align}
A={}&\sum_i \wh{m^2}_i=\Delta m^2_{21}+\Delta m^2_{31}+a\,,\\
B={}&\sum_{i<j}\wh{m^2}_i \wh{m^2}_j=\Delta m^2_{21}\Delta m^2_{31}\\
&+a\left[ \Delta m^2_{21}(1-|U_{e 2}|^2)+ \Delta m^2_{31}(1-|U_{e 3}|^2)\right]\,, \notag
\end{align}
and are also linear in $a$.
Everything to this point is exact except for the approximation as a two-flavor probability written in eq.~\ref{eq:NPZ probability}.

\begin{figure}
\centering
\includegraphics[width=\columnwidth]{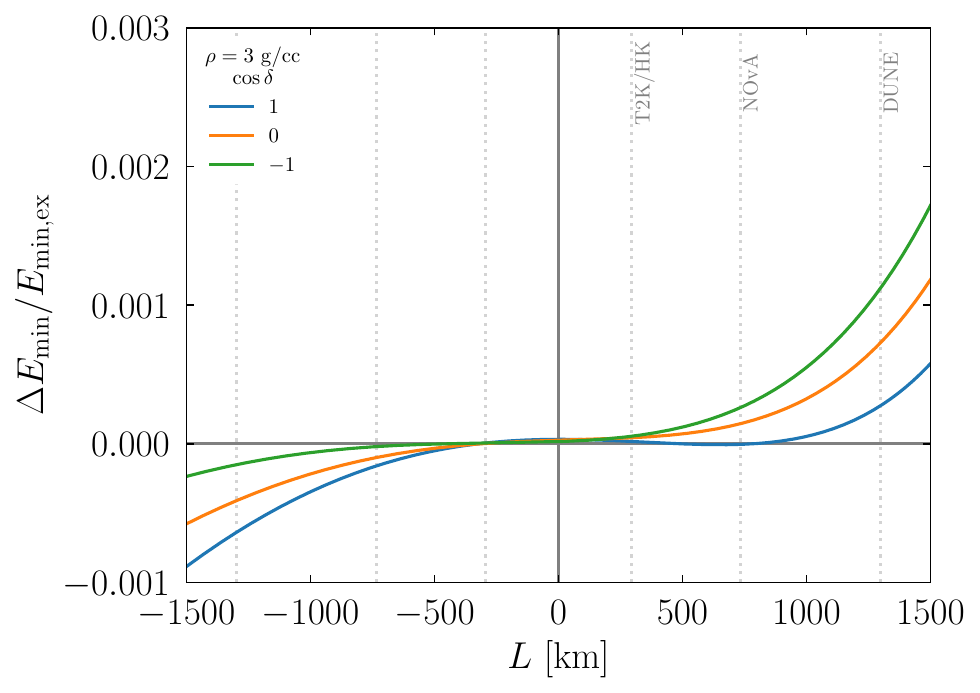}
\caption{The fractional error in the approximation to the minimum energy using eqs.~\ref{eq:Dmsqmm eigenvalues} and \ref{eq:NPZ probability} relative to the actual energy at the first oscillation minimum at $\rho=3$ g/cc and for different values of $\cos\delta$.
Antineutrinos are denoted by $L<0$.
The vertical dashed lines indicate the baselines of current and future experiments.}
\label{fig:min approx}
\end{figure}

\begin{figure*}
\centering
\includegraphics[width=\textwidth]{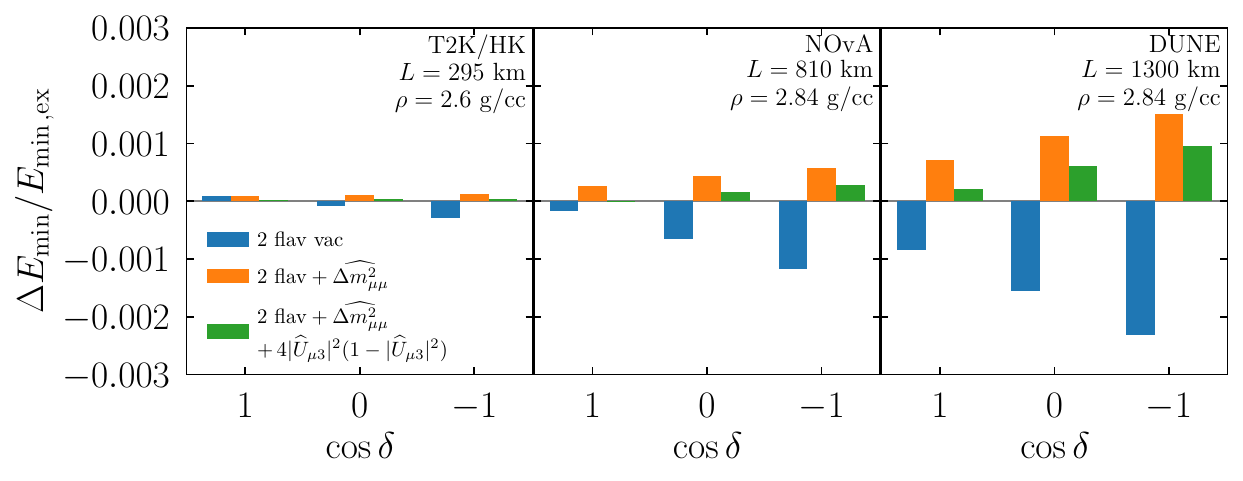}
\caption{The fractional error in the energy of the first minimum between the exact minimum and various approximations for T2K/HK (left), NOvA (middle) and DUNE (right) and at different values of $\cos\delta$.
The first approximation (blue) is the two-flavor vacuum approximation from \cite{Nunokawa:2005nx}, see eq.~\ref{eq:Dmsqmm effective}.
The second approximation (orange) uses the same structure to the probability, but the energy dependent frequency defined in eq.~\ref{eq:Dmsqmm eigenvalues}.
The third approximation (green) also uses the same structure and the energy dependent frequency, and also the energy dependent amplitude of the oscillation defined in eq.~\ref{eq:NPZ probability}.
In vacuum the two-flavor vacuum approximation has an accuracy for this quantity $\lesssim3\times 10^{-5}$.}
\label{fig:improvement}
\end{figure*}

We compare the energy at the minimum of the two-flavor probability in eq.~\ref{eq:NPZ probability} to the energy at the minimum of the exact three-flavor probability in fig.~\ref{fig:min approx} and find that the two-flavor approximation is accurate at the $<0.2\%$ level.
We also see that the remaining three-flavor correction accounts for a significant fraction, more than half, of the total contribution, by comparing this figure with fig.~\ref{fig:Emin vs L}.
We have also found that the most important correction to the very small error is addressed with elevating the frequency to that in matter: $\wh{\Delta m^2_{\mu\mu}}$ with a smaller contribution coming from the amplitude of the oscillation: $|\wh U_{\mu3}|^2(1-|\wh U_{\mu3}|^2)$.
The remaining error is due to the solar terms present in the full three-flavor expression.
This evolution of the error can be visually seen in fig.~\ref{fig:improvement} for different values of $\cos\delta$ and different experiments.
We have also checked and it is not possible to gain additional improvement without adding in additional $\wh{\Delta m^2_{21}}L/4E$ corrections.

Thus the only non-trivial dependence on energy and the matter effect in the approximation is 
$\wh{m^2_3}$.
While the exact expression for $\wh{m^2_3}$ is the solution to a cubic equation \cite{cardano} which is extremely messy and challenging to directly approximate, an approach through approximate diagonalization of the Hamiltonian does lead to excellent approximations for neutrino oscillations \cite{Minakata:2015gra,Denton:2016wmg}, see also \cite{Agarwalla:2013tza,Denton:2018fex}.
Specifically we consider this approximation to $\wh{m^2_3}$:
\begin{align}
\wh{m^2_3} &\approx\frac12\left[a +\Delta m^2_{ee}\left(1+ \sqrt{\left(1-\frac{a}{\Delta m^2_{ee}}\right)^2+ \frac{4s_{13}^2a}{\Delta m^2_{ee}}} \right) \right] \notag \\& \quad +s_{12}^2\Delta m^2_{21}\,,
\end{align}
where $\Delta m^2_{ee} \equiv c^2_{12}\Delta m^2_{31}+ s^2_{12}\Delta m^2_{32}$ whose sign is determined by the mass ordering.
This is an excellent approximation for $\wh{m^2_3}$; the fractional error is $<10^{-5}$ for long-baseline accelerator neutrinos and the maximum error of $<10^{-4}$ is near the atmospheric resonance.
If one substitutes this approximation into the two-flavor approximations for $\nu_\mu$ disappearance, there is no relevant deviation in the precision of the approximations.

To summarize this section, we have shown that using a two-flavor picture of $\nu_\mu$ disappearance is extremely accurate.
We have also shown the three-flavor corrections are somewhat less important then the matter effects as can be seen in the asymmetry figure, fig.~\ref{fig:asymmetry}, as well as in fig.~\ref{fig:improvement}.
The matter effects change both the frequency ($\wh{\Delta m^2}_{\mu\mu}$) and the amplitude ($4|\wh U_{\mu3}|^2(1-|\wh U_{\mu3}|^2)$) of the oscillations. However all of these effects are at the level of a few tenths of a percent or less for the current and proposed accelerator based experiments.
We have also found that further approximating the frequency using tools developed in past works retains the high level of precision necessary to understand the very small effects under consideration.

\section{Discussion}
\label{sec:discussion}
There are also other approximations in the literature aiming to understand the impact of the matter effect on $\nu_\mu$ disappearance which have their own advantages and disadvantages.
One of the earliest such approximations \cite{Arafune:1997hd} from 1997 takes the matter effect and $\Delta m^2_{21}$ to be small for a given experiment.
Near the oscillation minimum they write
\begin{multline}
P_{\mu\mu}\approx1-4c_{13}^2s_{23}^2\sin^2\frac{\Delta m^2_{31}L}{4E}\left[(1-c_{13}^2s_{23}^2)\right.\\
\left.-2\frac a{\Delta m^2_{31}}s_{13}^2(1-2c_{13}^2s_{23}^2)\right]\,,
\end{multline}
plus other terms that are zero near the minimum.
This approximation allows us to see that the role of the matter effect near the minimum is small for several reasons.
The quantity in square brackets is $[0.46+0.0033\frac a{\Delta m^2_{31}}]$ and $\frac a{\Delta m^2_{31}}\approx1/5$ for DUNE and smaller for the other long-baseline experiments.
So it is straightforward to see that this approximation says that the relative impact of the matter effect is at the $0.2\%$ level or less, consistent with our modern results.
We have numerically verified that the full version of the expression from \cite{Arafune:1997hd} gets the order of magnitude correct for the impact of the matter effect at the minimum for long-baseline experiments, although it is not extremely precise in reconstructing the features.

Another approximation for $\nu_\mu$ disappearance in matter appears in \cite{Akhmedov:2004ny} (eq.~33) which drops higher order terms\footnote{Note that this is technically only an expansion in $\Delta m^2_{21}/\Delta m^2_{31}$, not $s_{13}$; see \cite{Barenboim:2019pfp}.} in $\Delta m^2_{21}/\Delta m^2_{31}$ and $s_{13}$.
At seven lines long, however, it defies any realistic approximation or possibility to determine the relative role in the matter effect of any useful approximation.

More recently, a similar question to that of this paper was asked in \cite{Minakata:2017ahk}.
This study focused on the energy and baseline dependence that is necessarily present in the definition of $\wh{\Delta m^2_{\mu\mu}}$.
The expression presented there, however, diverges at $a \approx \Delta m^2_{31}$.
While this does not happen for long-baseline accelerator neutrinos, it does for atmospheric neutrinos at the atmospheric resonance, which provides some cause for concern.

Compared to the expected precision of upcoming experiments, the effects considered here are smaller than the expected precision, but likely not small enough to ignore.
The precision with which an experiment can determine $\Delta m^2_{31}$ (or equivalently $\Delta m^2_{32}$) is directly related to the determination of the energy of the oscillation minimum and the shape around the minimum.
DUNE and HK are expected to eventually reach a precision on $\Delta m^2_{32}$ of $\sim0.5\%$ \cite{DUNE:2020ypp,DUNE:2022aul,Hyper-Kamiokande:2018ofw}.
For DUNE this is still larger than the impact of the matter effect, but only by a factor of a few; for HK this is much larger than the impact of the matter effect.

In this article we have focused on $\nu_\mu$ disappearance for long-baseline accelerator neutrinos where the matter effect is very small.
In atmospheric neutrinos, the effect is larger making it harder to approximate the disappearance channel in atmospherics, see e.g.~\cite{Gandhi:2004md,Cogswell:2014kba}, due to both the longer baseline and higher density in the center of the Earth.
This larger matter effect provides some information about the mass ordering due to both a shifting of the location of the minimum as well as shape effects that can be described via the asymmetry parameter presented in eq.~\ref{eq:asymmetry} above.
A detailed discussion of the nature of these effects is beyond the scope of this paper, but is important for current and future atmospheric neutrino experiments.

\section{Conclusions}
\label{sec:conclusions}
The matter effect is intimately tied with the mass ordering and is an interesting phenomenon to study in its own right.
The $\nu_\mu$ disappearance channel is a well-studied channel with many expected improvements in the measurements in coming years.
In principle it would seem that the matter effect should play a small, but important, role in $\nu_\mu$ disappearance as it does for $\nu_e$ appearance.
Due to a number of subtle effects, however, the role of the matter effect is very small, $<0.3\%$ at DUNE for multiple potential observables.

A clear description of the smallness is challenging because three different effects all contribute comparably at the $\mathcal O(0.1\%)$ level to the deviation from the simple vacuum two-flavor picture:
\begin{enumerate}
\item $\wh{\Delta m^2_{\mu\mu}}(a)$: The biggest effect, barely, is on the frequency,
\item $4|\wh U_{\mu3}|^2(1-|\wh U_{\mu3}|^2)(a)$: The impact of the matter effect on the amplitude should also be considered,
\item The three-flavor correction coming in due to the $\Delta m^2_{21}$ oscillations is important as well.
\end{enumerate}
To understand the size of these effects given our current knowledge of the oscillation parameters, we have studied the energy at the oscillation minimum as well as a shape parameter, defined as an asymmetry in inverse energy.
These studies show that the matter effect appears at the level of $\lesssim0.3\%$ for DUNE and even lower for HK $\lesssim0.04\%$.
Thus while it is below the expected precision of upcoming experiments, it probably cannot be ignored at DUNE once they have acquired high-statistics.
That is, the matter effect \emph{can be ignored} for all long-baseline accelerator disappearance measurements except for when DUNE reaches their final high statistics stage, many years from now.
As the density of the Earth is often marginalized over in long-baseline studies, this simplifies numerical studies that can get very computationally expensive.

\begin{acknowledgments}
PBD acknowledges support by the United States Department of Energy under Grant Contract No.~DE-SC0012704.
PBD wishes to thank the Fermilab High Energy Theory Group for hospitality during the course of this project.
SJP acknowledges support by the United States Department of Energy under Grant Contract No.~DE-AC02-07CH11359.
SJP wishes to thank the Brookhaven High Energy Theory Group for hospitality during the course of this project.
This project has also received support from the European Union's Horizon 2020 research and innovation programme under the Marie  Sklodowska-Curie grant agreement No 860881-HIDDeN as well as under the Marie Skłodowska-Curie Staff Exchange grant agreement No 101086085 - ASYMMETRY.
\end{acknowledgments}

\appendix

\section{Biprobabilities}
\label{biprobabilities}

In vacuum, the dominant $\delta$ dependence in the probability for the three channels is
\begin{align}
\left[ 
\begin{array}{c}
 P_{\mu e}   \\[1mm]
 P_{\mu \tau} \\[1mm]
 P_{\mu \mu}
\end{array}
\right]& \sim 8 J_r  \sin \Delta_{21} \sin \Delta_{31} \notag
\\&  \hspace{-1cm} \left( \left[ 
\begin{array}{c}
1  \\[1mm]
2 s^2_{23}-1\\[1mm] 
-2 s^2_{23}
\end{array}
\right] \cos \Delta_{32} \cos \delta 
+\left[
\begin{array}{c}
-1  \\[1mm]
+1\\[1mm]
0
\end{array}
\right]  \sin \Delta_{32} \sin \delta \right)  \,.
\end{align}
Note that this $\delta$ dependence vanishes in the sum as it should since  $P_{\mu \mu}+P_{\mu e} +P_{\mu \tau} =1 $.  The coefficients of $ \sin^2 \Delta_{21 }$ in the probability also have $\cos \delta$ (and
 $\cos^2 \delta$ dependence for $P_{\mu \mu},\, P_{\mu \tau}$)  but  $ \sin^2 \Delta_{21 }$ is tiny for T2K, NOvA and DUNE.\\

For neutrinos and NO, at energies above 150 MeV (the solar resonance at 3 g/cc) and significantly below 12 GeV (the atmospheric resonance at 3 g/cc), the dominant terms in the oscillation probabilities are
\begin{align}
P_{\mu e} &\approx 4 |\wh{U}_{\mu 3}|^2 |\wh{U}_{e 3}|^2 \sin^2 \wh{\Delta}_{32}  \notag \\
P_{\mu \tau} &\approx 4 |\wh{U}_{\mu 3}|^2 |\wh{U}_{\tau 3}|^2 \sin^2 \wh{\Delta}_{31} \\
P_{\mu \mu} &\approx  1-4|\wh{U}_{\mu 3}|^2(1- |\wh{U}_{\mu 3}|^2)  \sin^2 \wh{\Delta}_{\mu \mu}. \notag
\end{align}
where $\wh{\Delta}_{ij} \equiv \Delta \wh{m^2}_{ij}L/(4E)$.
$  \Delta \wh{m^2}_{\mu \mu}$ is a weighted average of  $\Delta \wh{m^2}_{31}$ and $ \Delta \wh{m^2}_{32}$ defined in eq.~\ref{eq:Dmsqmm eigenvalues}.

The amplitudes of the oscillations in matter are given by
\begin{align}
|\wh{U}_{e 3}|^2& =  \wh{s^2}_{13} ,  \quad  |\wh{U}_{\mu 3}|^2\approx  \wh{c^2}_{13}  s^2_{23}, \quad  |\wh{U}_{\tau 3}|^2\approx  \wh{c^2}_{13} c^2_{23}\,,\\
\wh{s^2}_{13}  &\approx s^2_{13} + 2s^2_{13} c^2_{13} \frac{a}{\Delta m^2_{ee}}\,,
\end{align}
where the approximations in the first line are extremely precise because $\theta_{23}$ does not evolve much in matter.
These then combine to the relevant terms in the probabilities as
\begin{align}
4|\wh{U}_{\mu 3}|^2 |\wh{U}_{e 3}|^2 \approx{}& 4|{U}_{\mu 3}|^2 |{U}_{e 3}|^2  \\ &
+ 8|{U}_{\mu 3}|^2 |{U}_{e 3}|^2(1-2 |{U}_{e 3}|^2) \frac{a}{\Delta m^2_{ee}} \notag \, , \\
4|\wh{U}_{\mu 3}|^2 |\wh{U}_{\tau 3}|^2 \approx{}& 4|{U}_{\mu 3}|^2 |{U}_{\tau 3}|^2  \\ &
+ 8|{U}_{\mu 3}|^2 |{U}_{e 3}|^2(-2 |{U}_{\tau 3}|^2) \frac{a}{\Delta m^2_{ee}} \notag \, , \\
4|\wh{U}_{\mu 3}|^2(1- |\wh{U}_{\mu 3}|^2) \approx{}& 4|{U}_{\mu 3}|^2 (1-|{U}_{\mu 3}|^2 ) \\ &
- 8|{U}_{\mu 3}|^2 |{U}_{e 3}|^2(1-2 |{U}_{\mu 3}|^2) \frac{a}{\Delta m^2_{ee}}  \notag \, .
\end{align}
Since $2 |{U}_{\tau 3}|^2 \approx 1$ and since $|U_{e3}|^2\ll1$, the matter effect on the amplitude is approximately the same size and opposite in sign for $P_{\mu \tau}$ as for $P_{\mu e}$.  There is total cancellation  of this leading term in $P_{\mu \mu}$ , if $ |{U}_{\mu 3}|^2 = 1/2$.
The size of the shift for muon neutrino disappearance is
\begin{align}
8|{U}_{\mu 3}|^2 |{U}_{e 3}|^2(1-2 |{U}_{\mu 3}|^2) \frac{a}{\Delta m^2_{ee}} \sim 0.008  ~\left( \frac{E}{11{\rm\ GeV}} \right) \, .
\end{align}
This is roughly one tenth the size of the effect for the other channels.

The effect on the frequency of the oscillation is more complicated as matter effects  on the frequency of the oscillation are larger for  $\nu_\mu  \rightarrow \nu_e$ than for  $\nu_\mu  \rightarrow \nu_\tau$ and $\nu_\mu  \rightarrow \nu_\mu$. However the amplitude of the oscillations is much smaller for  $\nu_\mu  \rightarrow \nu_e$ than for the other channels.

\section{Hamiltonian}
\label{sec:hamiltonian}
Here we explicitly write out the Hamiltonian including the matter effect in the flavor basis which is generally useful, and is specifically useful for the quantity $m_\mu^2$ discussed in the main text,
$2EH_{\alpha \beta} =  \sum_i m^2_i U_{\alpha i}U^*_{\beta i}+a\delta_{\alpha e} \delta_{\beta e}  $, with $m^2_1=0$ and the PDG convention for U (using $J_{rr}\equiv s_{13}s_{12}c_{12}s_{23}c_{23}$, see e.g.~\cite{Jarlskog:1985ht,Denton:2016wmg}):
\begin{align}
2EH_{ee}={}&a+s_{13}^2\Dmsqee+s_{12}^2\Delta m^2_{21}\,,\\
2EH_{e\mu}={}&s_{13}c_{13}s_{23}e^{-i\delta}\Dmsqee+c_{13}s_{12}c_{12}c_{23}\Delta m^2_{21}\,,\\
2EH_{e\tau}={}&s_{13}c_{13}c_{23}\Dmsqee-c_{13}s_{12}c_{12}s_{23}e^{i\delta}\Delta m^2_{21}\,,\\
2EH_{\mu\mu}={}&c_{13}^2s_{23}^2\Dmsqee
+(c_{12}^2c_{23}^2+s_{12}^2s_{23}^2)\Delta m^2_{21} \, ,\\
&-2J_{rr}\cos\delta\Delta m^2_{21}\,,\\
2EH_{\mu\tau}={}&e^{i\delta}\left[s_{23}c_{23}(c_{13}^2\Dmsqee-(c^2_{12}-s^2_{12})
 \Delta m^2_{21})\right.\nonumber\\
&\left.-s_{13}s_{12}c_{12}(c_{23}^2e^{-i\delta}-s_{23}^2e^{i\delta})\Delta m^2_{21}\right]\\
2EH_{\tau\tau}={}&c_{13}^2c_{23}^2\Dmsqee+(s_{12}^2c_{23}^2+c_{12}^2s_{23}^2)\Delta m^2_{21}\nonumber\\
&+2J_{rr}\cos\delta\Delta m^2_{21} \, .
\end{align}
Then $H_{\alpha\beta}=H_{\beta\alpha}^*$.
The norm square of the elements of PMNS matrix in the PDG convention are:
\begin{align}
|U_{e3}|^2={}&s^2_{13}, \quad |U_{\mu3}|^2=c^2_{13}s^2_{23}, \quad  |U_{\tau 3}|^2=c^2_{13} c^2_{23}\, , \notag \\
|U_{e2}|^2={}&c^2_{13} s^2_{12}\, ,  \notag \\
|U_{\mu2}|^2={}&c_{12}^2c_{23}^2-2J_{rr} \cos\delta +s^2_{13}s_{12}^2s_{23}^2  \, ,\notag \\
|U_{\tau 2}|^2={}&c_{12}^2s_{23}^2+2J_{rr} \cos\delta +s^2_{13}s_{12}^2c_{23}^2 \notag \\
\end{align}
and the remaining elements can be best expressed by unitarity: $|U_{\alpha 1}|^2=1-|U_{\alpha 2}|^2-|U_{\alpha 3}|^2$ for $\alpha=e,\mu,\tau$.

\bibliography{Dmsqmm}

\end{document}